\documentclass[12pt]{iopart}
\usepackage[
  a4paper,               
  left=2.5cm,            
  right=2.5cm,
  bottom=3cm              
]{geometry}

\expandafter\let\csname equation*\endcsname=\relax 
\expandafter\let\csname endequation*\endcsname=\relax 
\usepackage{amsmath,amsfonts,amssymb}
\usepackage{float}

\usepackage[utf8]{inputenc}
\usepackage[english]{babel}
\usepackage[T1]{fontenc}
\usepackage{upgreek}
\usepackage{xcolor}

\usepackage[colorlinks=true, linkcolor=blue, citecolor=orange, filecolor=magenta, urlcolor=cyan]{hyperref}
\usepackage{graphicx}

\usepackage{csquotes}
\usepackage[backend=biber,style=numeric,sorting=none]{biblatex}
\usepackage[noabbrev,nameinlink,capitalise]{cleveref}
\crefname{appendix}{}{}
\usepackage{xspace}
\usepackage[font=footnotesize,labelfont=bf,margin={75pt,0pt}]{caption}
\usepackage{subcaption}
\usepackage{acronym}
\usepackage{placeins}

\newcommand{\nessai}{\texttt{nessai}\xspace}

\newcommand{\bayesbeat}{\texttt{bayesbeat}\xspace}

\acrodef{FFT}[FFT]{fast Fourier transform}
\acrodef{CTN}[CTN]{coating brownian thermal noise}
\acrodef{FEA}[FEA]{finite element analysis}
\acrodef{ESD}[ESD]{electrostatic drive}
\acrodef{SNR}[SNR]{signal-to-noise ratio}
\acrodef{GeNS}[GeNS]{Gentle Nodal Suspension}

\begin{document}

\newcommand{\aratio}{\rho}
\newcommand{\ascale}{A_\textrm{scale}}


\newcommand{\logBFModelOneModelOneStationaryThirtyOne}{4047.2}
\newcommand{\logBFModelThreeModelOneThirtyOne}{1943.4}
\newcommand{\logBFModelThreeModelOneStationaryThirtyOne}{5990.7}

\newcommand{\logBFModelThreeModelOneFourteen}{-0.1}
\newcommand{\logBFModelThreeModelOneStationaryFourteen}{-2.4}

\newcommand{\logBFModelOneModelOneStationaryRealDataOne}{4285.7}
\newcommand{\logBFModelThreeModelOneRealDataOne}{17.4}

\newcommand{\logBFModelOneModelOneStationaryRealDataFour}{4540.6}
\newcommand{\logBFModelThreeModelOneRealDataFour}{3.5}

\title[Bayesian Inference for Diagnosing Issues in Measurements of Disk Resonators]{Use of Bayesian Inference to Diagnose Issues in Experimental Measurements of Mechanical Disk Resonators}

\author{Simon C. Tait\textsuperscript{1,2}\footnote{\label{authorfootnote}A portion of the work for this paper was carried out while the author was at the University of Glasgow.}\addtocounter{footnote}{-1}\addtocounter{Hfootnote}{-1}, Michael J. Williams\textsuperscript{2,3}\footnotemark, Joseph Bayley\textsuperscript{1}, Bryan W. Barr\textsuperscript{1}, Iain Martin\textsuperscript{1}}

\address{\textsuperscript{1}California Institute of Technology, Pasadena, California, USA}
\address{\textsuperscript{2}School of Physics and Astronomy, University of Glasgow, Glasgow G12 8QQ, Scotland}
\address{\textsuperscript{3}Institute of Cosmology and Gravitation, University of Portsmouth, Portsmouth PO1 3FX, United Kingdom}

\ead{stait@caltech.edu}

\begin{abstract}

Gravitational wave detectors, such as LIGO, are predominantly limited by coating Brownian thermal noise (CTN), arising from mechanical losses in the Bragg mirror coatings used on test-mass optics. Accurately characterizing and minimizing these losses is crucial for enhancing detector sensitivity. This paper introduces a general mathematical and statistical framework leveraging Bayesian inference to precisely analyse mechanical ring-down measurements of disk resonators, a standard method for quantifying mechanical loss in coating materials.
Our approach presents a refined model that fully captures the non-linear behaviour of beam spot motion on split photodiode sensors, significantly improving upon traditional simplified exponential-decay methods. We achieve superior estimation accuracy for decay constants ($\tau_1$ and $\tau_2$), especially for measurements exhibiting larger oscillation amplitudes. Specifically, we observe improvements in estimation accuracy by up to 25$\%$ over traditional methods, with strong Bayesian evidence favouring our framework. Our simulations and experimental validations reveal that previously discarded measurements due to fitting inaccuracies can now be reliably analysed, maximizing the use of available data. This enhanced analytical capability not only provides more precise mechanical loss estimations but also offers deeper insights into systematic issues affecting disk resonator measurements, paving the way toward improved coating materials and ultimately, more sensitive gravitational wave detectors.

\end{abstract}

{\it Keywords}: {Mechanical Loss,  Bayesian Inference, Nested Sampling, GeNS} 

\submitto{Classical and Quantum Gravity}

\maketitle

\section{Introduction}
\label{introduction}

Gravitational wave astronomy has transformed our understanding of the cosmos by allowing us to observe ripples in spacetime from cataclysmic events. 
Since the landmark detection of GW150914\_095045 in 2015~\cite{Abbott2016} and over 200+ subsequent detections~\cite{Abbott_2023}, we have gained unprecedented insights into black holes and neutron stars, underscoring the importance of enhancing the sensitivity of gravitational wave detectors.
The sensitivity of the LIGO \cite{LIGOScientific:2014pky}, detectors is currently limited by \ac{CTN} in the 100\,Hz\,-1000\,Hz range\,\cite{GW150914FirstDiscoveries} and significant efforts are being made to investigate new coating materials to reduce the impact of CTN on detector sensitivity, as well as reducing their impacts on other active detectors such as VIRGO~\cite{VIRGO:2014yos} and KAGRA~\cite{KAGRA:2020tym} .

These reflective coatings are used in the Fabry-Pérot cavities of the detector arms and consist of alternating layers of high (n$_{\text{H}}$) and low refractive index (n$_{\text{L}}$) oxide glasses. New low mechanical loss materials, such as titania doped germania, (Ti:GeO$_{x}$)~\cite{PhysRevLett.127.071101}, titania doped silica  (Ti:SiO$_{2}$) ~\cite{PhysRevLett.131.171401}, and amorphous silicon (aSi)~\cite{PhysRevLett.125.011102}, are being investigated as potential replacements for the high refractive index titania doped tantala (Ti:Ta$_{2}$O$_{5}$) currently used as these materials currently contribute to the  total level of CTN in the detector. There has been a concerted effort to understand these effects in the past ~\cite{Hong2013BrownianMirrors,Harry2018ThermalCoatings,Rowan2005ThermalDetectors,Yu1998InternalApproach} \textit{et al}. Each of these  has highlighted that any means of characterising the mecahncial loss responsible for this phenomenon  is crucial in understanding its effects on gravitational wave detector sensitivity.

\begin{figure}[H]
\centering
\includegraphics[width=0.6\linewidth]{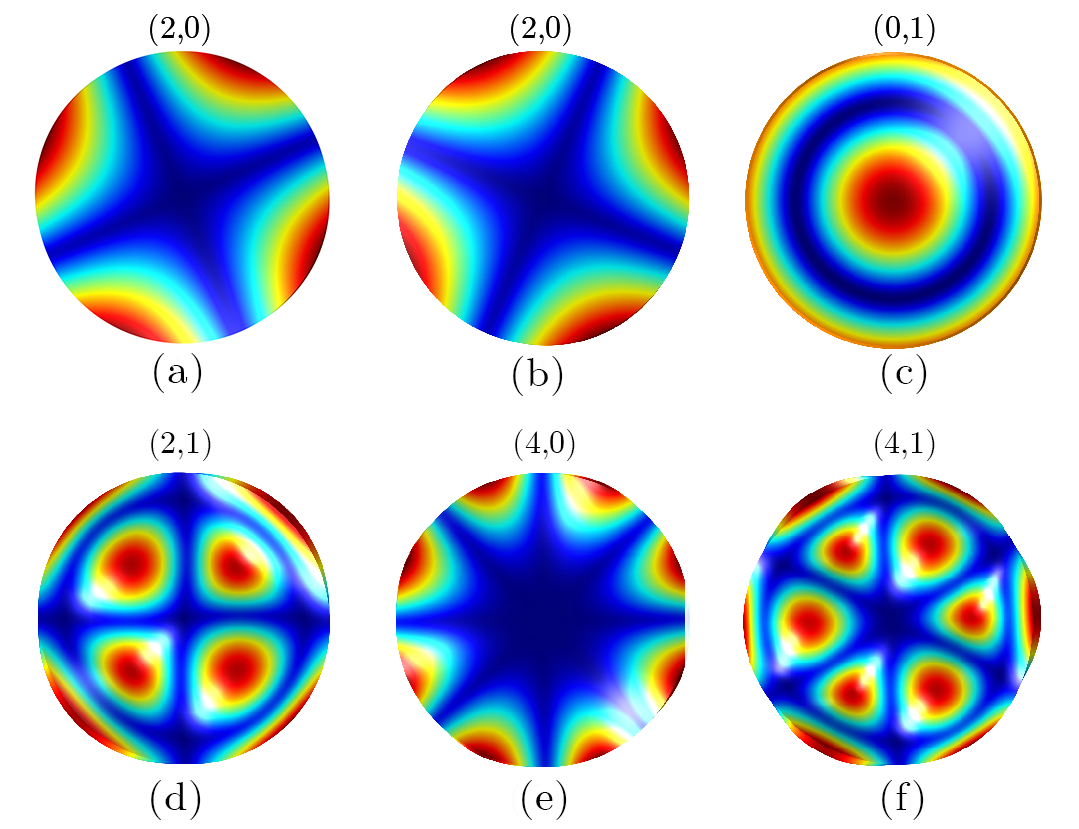}
\caption[Example of low frequency surface deformations of mode-shapes of a cylinder calculated with finite element modelling (COMSOL) listed in ascending frequency.]{Example of low frequency surface deformations of mode-shapes of a cylinder calculated with finite element modelling (COMSOL) listed in ascending frequency. Regions of large displacement are denoted in red, contrasting regions with little/no displacement, are shown in blue.}
\label{fig:cylinder_modes}
\end{figure}

Mechanical loss in coating materials, a property directly related to \ac{CTN}, can be determined by observing the decay of the resonant mechanical modes of a disk resonator before and after a coating material is applied. The mechanical modes of disk resonators are characterized by the number of nodal lines (symmetry line across the flat surface of the cylinder) and nodal circles (an unbroken radial `ring showing minimal deformation), denoted as ($m$,$n$) respectively~\cite{Tomassini2017}. Where  `$n$' is sometimes referred to as the  `mode family'.
The mechanical loss, and therefore the \ac{CTN}, is measured in a frequency range closely tied to the most sensitive range in the LIGO sensitivity band (100\,Hz-1000\,Hz) and the substrates are made of  isotropic SiO$_{2}$.  
 
From \ac{FEA} modelling of these samples with COMSOL Multiphysics~\cite{Comsol} the expected deformation and frequencies of these modes were calculated (see \cref{fig:cylinder_modes}). to the isotropy of the substrate, all mode-shapes where $m > 2$ \cref{fig:cylinder_modes}\textbf{(a,d-f)} will produce mode-pairs or mode-doublets at nearly the same frequency ($\Delta f \le 0.1\,\textrm{Hz}$). 
The deformation and frequency of these modes are identical, but rotated at $90/m$,  around a geometric reference frame with respect to the other as shown in \cref{fig:cylinder_modes}\textbf{(a)} and \cref{fig:cylinder_modes}\textbf{(b)}.
Experimentally, the exact frequency, deformation and mechanical loss of each of these modes in a pair  can differ slightly due to various factors such as ; geometric imperfections, stresses induced by thermal expansion coefficients during coating, and other effects~\cite{PhysRevD.70.082003}. 
Typically, each mode-pair is usually separated by $\le 0.5\,\textrm{Hz}$.

 Since these modes are close in frequency, they cannot be analyzed independently. Instead, the decay constants $\tau_{\text{1}}$ and $\tau_{\text{2}}$ for each mode in the mode-pair must be extracted simultaneously by exciting both modes together and observing the time-dependent surface deflection as the system undergoes free decay (or ringdown). The mechanical loss $\phi(f)$ or Quality Factor ($Q$) of a resonator is then determined by fitting this decay to an appropriate model, yielding 
\begin{equation}\label{eq:mechanical_loss}
    \phi(f_n) = \frac{1}{Q_n} = \frac{1}{\pi f \tau_{n}}.
\end{equation}
Precisely disentangling $\tau_{\text{1}}$ and $\tau_{\text{2}}$ is particularly challenging, their close proximity in frequency makes it difficult to separate their contributions to the observed decay. This ambiguity in extracting the decay constants directly impacts the accuracy of the mechanical loss estimation and, consequently, the inferred level of \ac{CTN} in a gravitational wave detector. The conventional approach, which involves fitting a simple model of exponentially decaying sinusoids, often struggles to resolve the two decay rates, leading to poor fits and, in some cases, complete failure of the fitting process. 

 In this work, we present a general, comprehensive framework for improving the precision of mechanical loss estimates when measured using the \ac{GeNS} technique~\cite{Cesarini2009AMaterials} described in \cref{sec:experimental_setup}. This method is designed to account for potential variations between different sets of apparatus, without the need for extensive calibration.  
 We revisit a simpler mathematical model previously used to estimate these parameters from the vibrational modes of a perfect cylinder, then expand upon this to account for non-linearities in coupled mode-pair measurements. 
 Furthermore, we develop a more complete model of photodetector outputs to enhance simulation efficiency and accuracy, using Bayesian inference to improve the reliability of parameter estimation. 
 
In \cref{sec:experimental_setup}, we introduce the experimental setup and data collection methods, in ~\cref{sec:models} we will introduce and explore a novel analytical model designed to accurately characterize the signals detected during decay measurements from a resonating sample using a split photodiode. 
In the \cref{sec:bayes}, we describe Bayesian statistics and \nessai\,\cite{Williams_2021}, a Nested Sampling algorithm that utilizes normalizing flows to accelerate sampling. 
In \cref{sec:simulated_data}, we validate the proposed noise and signal models using simulated data which we then use to analyse real data in \cref{sec:results}.

\section{Experimental Setup }
\label{sec:experimental_setup}

The mode-pairs of the disk resonator are measured using the \ac{GeNS} technique where each mode-pair is excited using an \ac{ESD}, then the surface distortion of the resonator is measured by reflecting a laser off the surface and sensing deflections in the alignment of the reflected beam on a split photodetector. For room temperature applications, as discussed in this paper; the sample substrate is typically made of fused SiO2 (76.2\,mm in diameter $\times$\,2.5\,mm thick). These samples are measured `uncoated' before having additional material layer(s) deposited on one, or both circular faces. This process allows the mechanical loss of the additional material(s) to be extracted. In order to identify the mode shape and expected frequencies of the sample under test, the surface deformations of various modes of a thin cylinder were determined through \ac{FEA} using COMSOL MultiPhysics \cite{Comsol}, which can be correlated to the frequencies of the measured sample. 
In a \ac{GeNS} set-up, the disk is supported at its centre of mass, therefore any mode shape which has minimal displacement in the centre (\textit{m\,$\ge$}0,1, see \cref{fig:cylinder_modes} will also experience minimal frictional damping. 
All measurements are carried out under high vacuum ($\le$\,1$\times$10$^{-5}$\,mbar) to reduce external gas damping effects.

\begin{figure}
    
    \centering
    \includegraphics[width=0.7\linewidth]{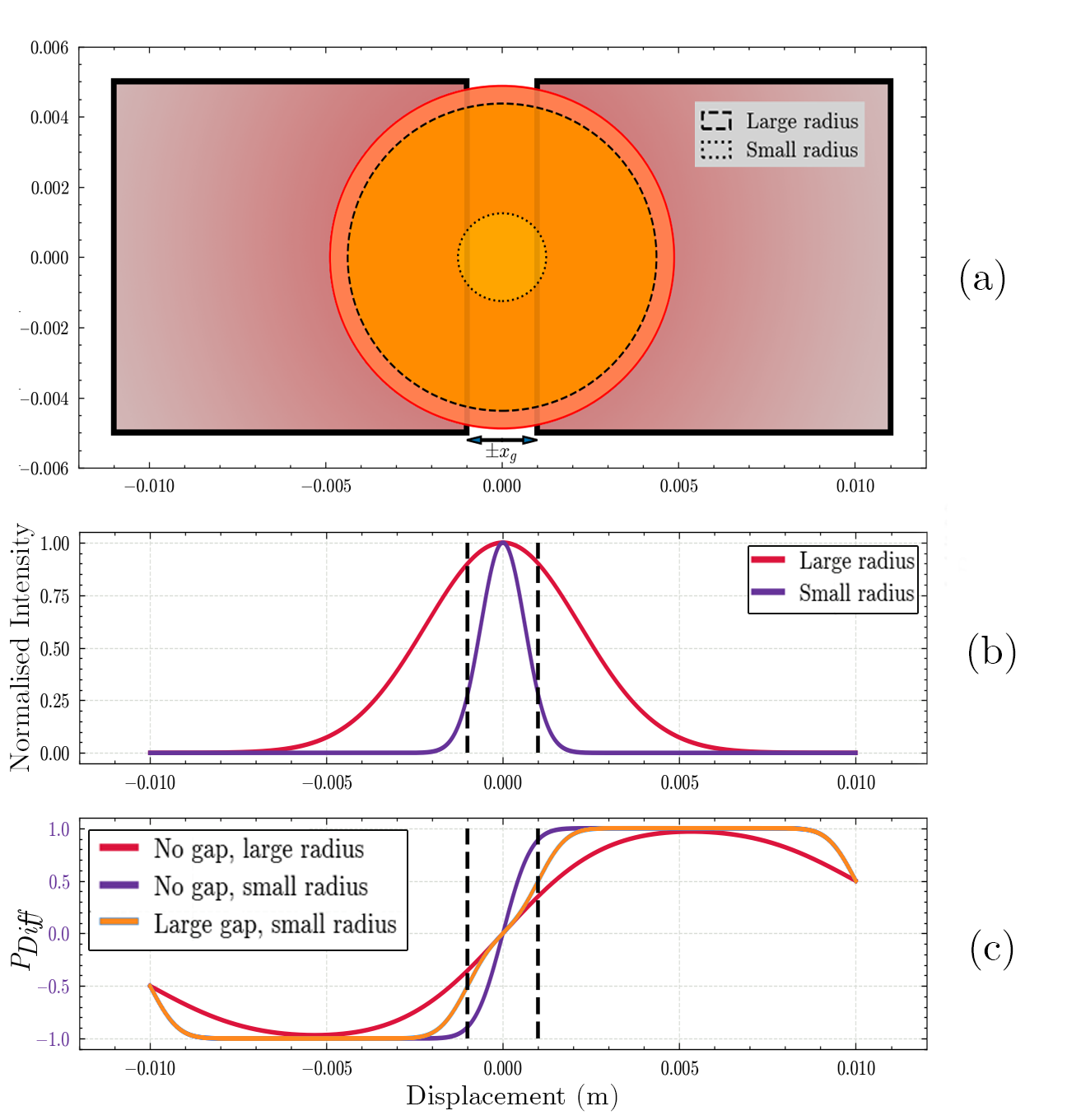}
    \caption{Simple illustration of a typical photodetector setup and non-linear signal behaviour. Top: Exaggerated example of two beams (small radius and large radius) centred on a split photodetector. Middle: normalised intensity profile of two smaller and larger radius beams incident on the photodetector and gap. Bottom: Scan of the two beams across the photodetector. The larger radius beam across a configuration with no gap (red) is non-linear and is the result of a Gaussian profile beam. The smaller beam on the same no-gap setup (purple) has the same shape but compressed around the origin indicating smaller beams mean more nonlinearity present for smaller displacements. And the smaller beam across the gap (orange) has a very clear chicane feature around the origin, indicating additional non-linear behaviour in the presence of a gap where smaller gaps give less non-linearity.}
    \label{fig:PD_Gap}
\end{figure}

 The laser beam is read out using a simple optical lever arrangement with a $5\,$mW helium neon (HeNe) laser reflecting off the oscillating surface of the sample, and centred on a split silicon photodetector. In this work the authors operate under the assumption of a split photodiode where the laser's trajectory is perpendicular to the sensor plane and strikes the photodiode at normal incidence. An illustration of this is shown in \cref{fig:PD_Gap}(a) where it is shown  that there is a non-zero sized gap between the two sides of the detector and that a Gaussian beam scanning across two linear photodiodes will intrinsically produce a non-linear response (shown in \cref{fig:PD_Gap}(b)). The output from this photodetector is the differential voltage \(V_D\) from the diffrence in power hitting each side of the photodiode ($\textrm{PD}_{L}$ and $\textrm{PD}_{R}$).

 To excite a given mode pair, LabView software is employed to first generate a swept sine wave that is passed to the \ac{ESD}; the frequency is slowly scanned up until a resonance signal is observed on $V_D$. The excitation is then stopped and $V_D$ is sampled at a rate of 250\,kS/s (250\,kHz), significantly above the Nyquist frequency of the highest mechanical mode observed in this setup (approximately 31\,kHz). Every 0.2\,s (50\,kS), a \ac{FFT} is taken, with the peak intensity in each iteration corresponding to a single data point in the recorded ringdown data \(d(t_i)\). This data can be defined by this periodigram expression 
\begin{equation}
    \label{sec:experimental:data}
    d(t_i) = \max \left\{ \left|\text{FFT}\left(V_D(t^{'}) \right) \right|^2 : t^{'} \in [t_i, t_i+\Delta t]  \right\},
\end{equation}
where \(V_D(t_i)\) is the voltage output from the detector, \(t_i\) is start time of segment \(i\) and \(\Delta t\) is the window over which we take the \ac{FFT}. This method allows one to directly visualise the decay in the sample's motion amplitude over time, and fitting such a decay as a function of time gives an estimate of the two different decay parameters $\tau_{\text{1}}$ and $\tau_{\text{2}}$, from which their mechanical loss can be calculated.

 Due to the varying decay durations across different samples, each with distinct material properties and decay constants, the sampling frequency was chosen to balance acquisition speed and resolution. While this ensures shorter decays are captured, it results in data bins wider than the mode-pair's separation frequency ($\Delta f < 0.5\,\textrm{Hz}$). 
 As a result, $d(t_i)$ represents a beat between two non-linearly decaying sinusoids, making the accuracy of the extracted parameters entirely dependent on the validity of the chosen model. Meaning any inaccuracies in capturing the true decay in amplitude; and consequently energy dissipation will directly affect the estimation of mechanical loss. Typically, data produced in this manner would be assumed by the summation of two decaying sinusoids \textit{A} and \textit{B}, from which the decay parameters could be extracted. While providing a foundational understanding of cylinder vibrations, it exhibits significant limitations when applied to complex scenarios, particularly in capturing higher-order/non-linear effects. A key shortcoming arises from its inability to account for the effects of laser motion on the optical readout.

\begin{figure}
    \centering
    \includegraphics[width=\linewidth]{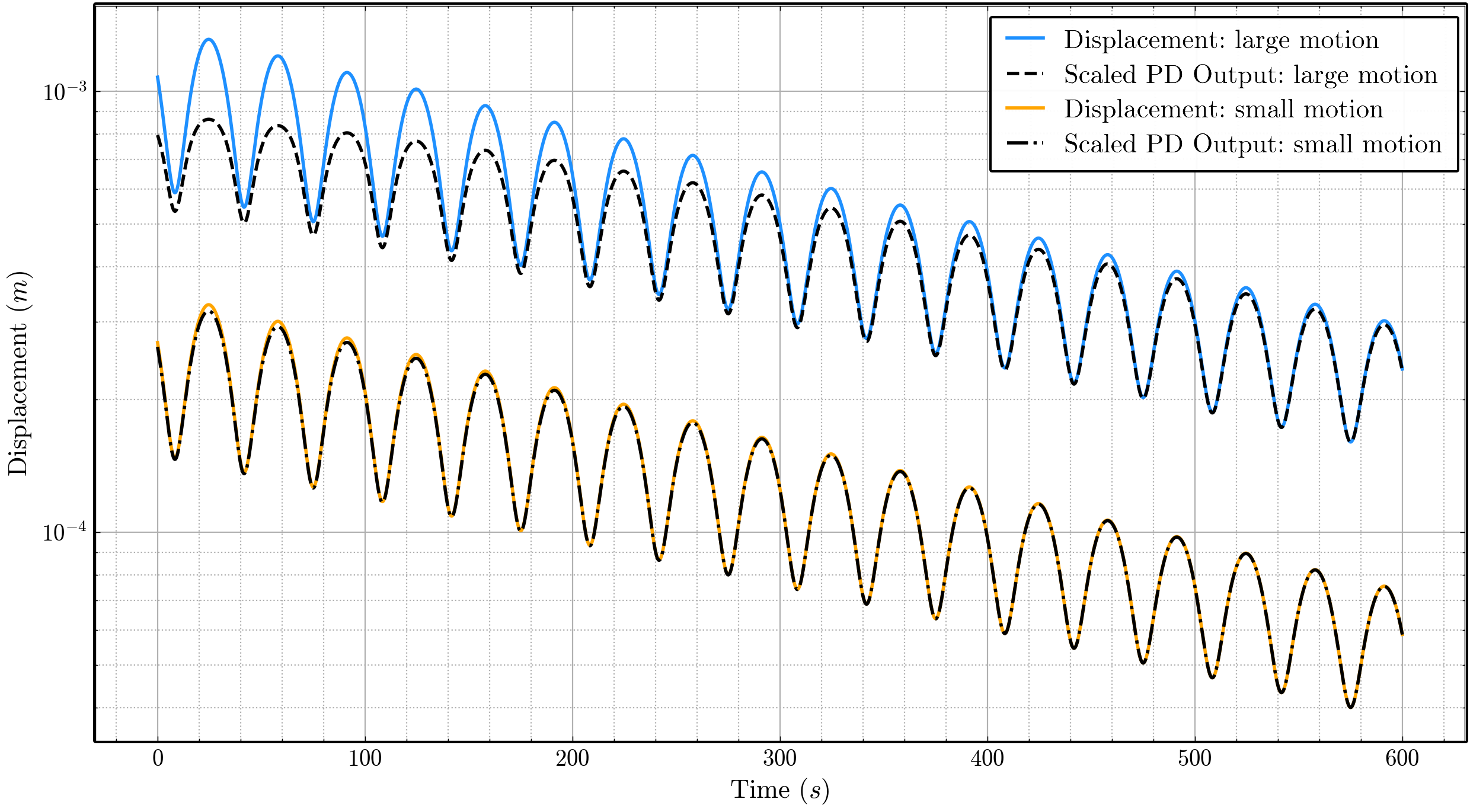}
    \caption{Simulated example of two exponential decay-pairs using the small beam spot in ~\cref{fig:PD_Gap}. The upper high displacement trace (initial mode-pair amplitudes of $1$\,mm and $0.4\,$mm) has clear distortion in the observed photodiode signal whereas in the lower low displacement trace (initial mode-pair amplitudes of $0.25\,$mm and $0.1$\,mm) the photodiode signal is much closer to the input displacement. It is clear from these traces that the exponential ringdowns are affected by amplitude dependant distortion and that a simple exponential fit would be insufficient to accurately extract the decay parameters of the mode-pairs, particularly for measurements where a stronger displacement is observed. }
    \label{fig:Exponential_decay_distortion}

\end{figure}

This issue is highlighted in \cref{fig:Exponential_decay_distortion}, where a comparison between the displacement of the beam position on the photodiode and the $d(t_i)$ readout can be seen with clear amplitude dependant distortion of the readout. While it is most obvious in the exaggerated high displacement example, the smaller displacement example does exhibit distortion, underscoring the need for a more comprehensive model capable of integrating higher-order interactions and the subtleties of laser movement and its interaction with the split photodiode sensor.

\section{Modelling the experiment}
\label{sec:models}

In this section, we present the mathematical models used to represent the experimental setup outlined in ~\cref{sec:experimental_setup}. We introduce three models of increasing complexity, referred to as $M_1$, $M_2$, and $M_3$ for clarity.

First, \cref{models:model1} describes a previously established model in which the output signal is represented as exponential decays of two oscillatory modes. Next, \cref{models:model2} extends this by incorporating a more detailed numerical model that accounts for laser intensity variations, the photodiode gap, the propagation of these effects to the measured output, and a more accurate representation of noise. Finally, \cref{models:model3} introduces an analytic approximation of the numerical model in \cref{models:model2}, providing a computationally efficient alternative while retaining key physical features.

\label{models:model1}
To further our understanding of disk resonator dynamics, we delve into a previously established model by \citeauthor{Vajente_high_throughput}~\cite{Vajente_high_throughput} that interprets the observed vibrational dissipation of  cylindrical samples. This model posits that the observed signal is a superposition of exponential decays from two oscillatory modes. We will assume that the measured signal arises from a pair of exponential decays from two modes in a degenerate pair, \textit{A} and \textit{B}
\begin{equation}
    A\,=A_{1} e^{- \frac{t}{\tau_{1}}} \sin{\left(\omega_{1} t + \varphi_{1} \right)} ,\ \ \ \
    B\,= A_{2} e^{- \frac{t}{\tau_{2}}} \sin{\left(\omega_{2} t + \varphi_{2} \right)}.
\label{eq:fancy_fitting_equation_waves}
\end{equation}
where $\varphi$\textsubscript{1} and $\varphi$\textsubscript{2} are the phases of each wave, the time-dependent phase is given by $\varphi$\textsubscript{2} = $\Delta\varphi$ + $\varphi$\textsubscript{1}, and $\tau$\textsubscript{1}, $\tau$\textsubscript{2} are the corresponding decay constants. The amplitudes \textit{A}\textsubscript{1} and \textit{A}\textsubscript{2} are assumed to lie within the upper and lower envelope of the curve at $t(0)$. It is assumed that each decaying wave contains information about the other, where the angular frequency of the oscillation \text{$\Delta\omega$} is expressed in terms of the beat frequency $f_{\text{beat}}$ with \text{$f_{\text{beat}}\,=\,f_{2}-f_{1}$}:
%
%
\begin{equation}
    \Delta\omega\,=\,2\pi \,f_{\text{beat}}.
    \label{eq:fancy_fitting_equation_omegadef}
\end{equation}
These equations then give two components, a high and a low frequency oscillating component. Adding both of these components in quadrature gives the intensity of the decaying waveform, allowing a simple formulation of the interference between both waves to be obtained:
\begin{equation}
\centering
    s_1^2(t) = I^2\,=\,{|A|^2 + |B|^2 + 2 |A| |B| \cos(\Delta\omega t + \Delta\varphi).}
\label{eq:model_1_equation}
\end{equation}


\subsection{\texorpdfstring{\textbf{Model 2 \((M_2)\):}}{Model 2 (M2):} Mathematical Description of Laser Intensity Calculations}
\label{models:model2}

In order to more accurately reflect the non-linearities observed in the real world data, a mathematical model of the split photodiode system was developed. This model incorporates the physical parameters of the experimental setup, such as the size of the Gaussian beam ($\sigma$) the gap between the photodiodes ($x_\textrm{g}$), and the dimensions of the sensors themselves. By integrating these parameters, the model offers a direct simulation of the experimental apparatus.

 Fundamentally, the decaying sinusoidal modes are measured by reflecting a laser beam off the surface of the sample under test and observing the deflection of the reflected beam on a split photodetector. The probe beam is considered to have a circular Gaussian profile ($\text{TEM}_{00}$) which for simplicity of calculation can be considered to have a normalised intensity ($I_n$) and a beam radius of $\omega_r = 2\sigma$. Thus the changing intensity along a plane $x$ for this Gaussian beam at a mean beam spot position of $\mu$ can be defined at any point on the axis of propagation as
\begin{equation}
I(x,\sigma, \mu) = I_n \exp\left(-\frac{1}{2}\left({\frac{x - \mu}{\sigma}}\right)^2\right).
\end{equation}
When the beam illuminates a split photodetector a change in the beam centre position ($\mu$) will result in a change in the power measured on the left and right sides of the detector. The power on the right-hand side photodiode \(I_{R}\) with edges at $x_\textrm{g}$ and $x_\textrm{e}$ (the gap between the photodiodes and the edge of the photodiode respectively) can be expressed as
\begin{equation}
    I_{R}(\sigma, \mu) = \int_{x_\textrm{g}}^{x_\textrm{e}} \frac{1}{\sigma \sqrt{2\pi}} \exp\left({-\left(\frac{x - \mu}{\sqrt{2}\sigma}\right)^2}\right) \textrm{d}x,
\end{equation}
then rearranging in terms of an error function
\begin{equation}
    \textrm{erf}(z) = \frac{2}{\sqrt{\pi}}\int_{0}^{z} e^{-t^2} \textrm{d}t,
\label{eq:errorFunction_simple}
\end{equation}
the equation becomes
\begin{equation}\label{model1:power:right}
    I_{R}(\sigma, \mu) =  \frac{1}{2} \left(\textrm{erf}\left(\frac{x_\textrm{e} - \mu}{\sqrt{2}\sigma}\right) - \textrm{erf}\left(\frac{x_\textrm{g} - \mu}{\sqrt{2}\sigma}\right)\right)
\end{equation}
and the power on the left side with edges $-x_\textrm{e}$ and $-x_\textrm{g}$ is given by 
\begin{equation}\label{model1:power:left}
    I_{L}(\sigma, \mu) =  \frac{1}{2} \left(\textrm{erf}\left(\frac{-x_\textrm{g} - \mu}{\sqrt{2}\sigma}\right) - \textrm{erf}\left(\frac{-x_\textrm{e} - \mu}{\sqrt{2}\sigma}\right)\right).
\end{equation}

 Error functions are mathematically complex and lack an analytical solution in terms of elementary functions, but they can be easily solved numerically. 
The simplest practical approach to model the recorded data is to numerically model the voltage produced by the detector.
The sinusoidal motion of the beam is defined by
\begin{equation}\label{eq:model2_mu_eqn}
    \mu = A + B + x_0,
\end{equation}
where \(A\) and \(B\) are defined in \cref{eq:fancy_fitting_equation_waves} and \(x_0\) is an arbitrary offset of the sinusoidal motion from the center of the split detector. 
The detectors output \(V_{S}\) can then be found by taking the difference between the photodiodes output defined by 
\begin{equation}
    V_{S}(\sigma, \mu) \propto I_R(\sigma, \mu) - I_L(\sigma, \mu).
\end{equation}

To obtain the resultant signal at the output \(s(t)\), the same process defined in \cref{sec:experimental_setup} is followed by splitting the timeseries measurements of \(V_S\) into multiple time segments, computing the \ac{FFT} for each segment and taking the maximum of its power to give 
\begin{equation}
\label{eq:model_2_equation}
s_2^2(t) = \max \left\{ \left|\text{FFT}\left(V_S\left(\sigma, \mu(t^{'})\right) \right) \right|^2 : t^{'} \in [t_i, t_i+\Delta t]  \right\}.
\end{equation}
For high \ac{SNR} signals the power in the frequency bin containing the mode-pair frequency is always the largest, therefore by taking the maximum of the power spectrum the amplitude of the beat at the mode frequency $(\omega_1 + \omega_2) / 2$ is found. 

\subsection{\texorpdfstring{\textbf{Model 3 \((M_3)\):}}{Model 3 (M3):} Analytic approximation}\label{models:model3}

In order to effectively utilise Nested Sampling algorithms described in \cref{subsec:nested} it is helpful to speed up the computation of the model described in \cref{models:model2} as the higher accuracy led to a more computationally expensive model. The first step is to simplify the problem.

In the experimental apparatus, we measure the beam radius ($\sigma \sim 1.5 \, \text{mm}$), which is small compared to the size of the photodetector ($x_\textrm{e} = 10 \, \text{mm}$), but large compared to the gap between the photodiodes ($x_\textrm{g} = 0.25 \, \text{mm}$) and the actual motion of the beam, which is on the order of a few hundred microns.
Any motion of the beam from the mean position where $|x - \mu| > 4\sigma$ results in only small changes in $I_L$ and $I_R$, because the motion of the beam is fully captured by the sensor. 
Therefore, the terms containing \(x_\textrm{e}\) in  \cref{model1:power:right,model1:power:left} can be set to be constant making the difference between the photodiodes power
\begin{equation}
V_{S}(\sigma, \mu) \propto  -\frac{1}{2} \left(\textrm{erf}\left(\frac{x_\textrm{g} - \mu}{\sqrt{2}\sigma}\right) + \textrm{erf}\left(\frac{-x_\textrm{g} - \mu}{\sqrt{2}\sigma}\right)\right).
\end{equation}

Given the amplitude dependent distortion in the simple exponential decay in~\cref{fig:Exponential_decay_distortion} it is clear that to accurately model the decay we must include non-linearities.
In order to investigate this we can take the Taylor expansion of the error function from~\cref{eq:errorFunction_simple}
\begin{equation}
\textrm{er}f(z) \approx \frac{2}{\sqrt{\pi}}\sum_{n=0}^{\infty}\frac{(-1)^nz^{2n+1}}{n!(2n+1)},
\end{equation}
and express $V_{S}$ up to some arbitrary number of terms T in a generalised power series form as
\begin{equation}\label{eq:power_series}
\begin{split}
    V_{S,T}(\sigma,\mu) & \approx  \sum_{k=0}^{T} C_k\mu^k\\
    & \approx C_0 + C_1\mu + C_2\mu^2 + \cdots + C_{T}\mu^{T},
\end{split}
\end{equation}
where $C_k$ are constant coefficients defined in terms of the fixed parameters \(x_\textrm{g}\) and \(\sigma\).
The nonlinear effect is immediately apparent if we consider a simple decaying sinusoidal motion with $\mu = A_1 e^{-\frac{t}{\tau_1}}\sin{(\omega_1 t)}$ and expand to $T=3$ to give

\begin{equation}
\begin{split}
V_{S,3} \approx &  \frac{3}{4}A_1^3 e^{-3\frac{t}{\tau_1}}\sin{(3\omega_1 t)} \left(\frac{\sqrt{2} }{6 \sqrt{\pi} \sigma^{3}}\right) \\
& + A_1 e^{-\frac{t}{\tau_1}}\sin{(\omega_1 t)} \left(\frac{\sqrt{2}}{\sqrt{\pi} \sigma} - \frac{\sqrt{2} x_{g}^{2}}{2 \sqrt{\pi} \sigma^{3}}\right)  \\
& - \frac{3}{4}A_1^3 e^{-3\frac{t}{\tau_1}}\sin{(\omega_1 t)} \left(\frac{\sqrt{2} }{6 \sqrt{\pi} \sigma^{3}}\right).
\end{split}
\label{eq:simple_3_term}
\end{equation}

The \(3\omega_1\) component clearly arises due to power in the \(\omega_1\) component shifting to the third harmonic. Additionally, the \(3\omega_1\) term decays more rapidly than the term at the fundamental frequency, and thus is more significant at the start of the decay where the amplitude of motion is larger.
Further expanding \(V_{S,T}\) to include more \(T\) terms will result in increased distortion at higher harmonics, up to a maximum frequency of \(T\omega_1\). It should be noted that with typical experimental conditions where \(A_1 << 1\) and \(x_\textrm{g}\ < \sigma\) the higher-order terms remain small compared to the fundamental, resulting in distortions of the \(\omega_1\) term rather than dominating the signal.

To achieve a complete model analytically, \(\mu\) is defined to have the same form as in \cref{eq:model2_mu_eqn} and we can apply this to \(V_{S,T}\).
Since the main carrier frequencies of oscillation are known from the experimental data, it is possible to perform an ``\ac{FFT}-like'' analysis in the frequency domain on only the one mode-pair of interest rather than calculating the whole spectrum from a time domain data series.
The value in an \ac{FFT} bin is a measure of the total power of all the signals that fall in that bin. Thus it is necessary to isolate the signals at frequencies within the relevant bin and determine the power of the combined beats of these signals. The simplest way to do this is to demodulate the signal at the mid frequency of the two sinusoids $\omega_p = \frac{\omega_1 + \omega_2}{2}$ (i.e. multiply the signal by a unitary amplitude sine wave at $\omega_p$) and then low-pass filter the result to only include terms inside the measurement bandwidth (i.e. the width of the \ac{FFT} bin).

We consider both the in-phase and quadrature-phase components of the original driving signal. Although this approach is more general than necessary for the current symmetric laser and photodiode setup, it will be important if the analysis is later expanded to account for any asymmetries in the beam shape or photodiode response. The signal we expect to measure \(s_3(t)\) is then the result of demodulating the \(V_{S, T}\) for both the in phase and quadrature phase components such that  
\begin{equation}
    \label{eq:model3_equation}
    s_3^2(t, \theta) = \left| V_{S, T}( \sigma, \mu(t)) \sin{(\omega_p t)} \right|^2 + \left| V_{D, T}(\sigma, \mu(t)) \cos{(\omega_p t)} \right|^2,
\end{equation}
 where \(\theta\) is a combination of all the model parameters. In the case where \(\mu\) is defined as in \cref{eq:model2_mu_eqn} the \(\theta : \left\{ \sigma, T, A_1, A_2, \omega_1, \omega_2, \tau_1, \tau_2, \varphi_1, \varphi_2 \right\}\). In this model low pass filtering is handled simply by discarding the terms which have a frequency greater than \(T(\omega_1 - \omega_2)\). Due to the increasing complexity of these equations when more terms are included, we do not show them in full here but they can be found in \bayesbeat~\cite{michael_j_williams_2024_12804193}. 
The overall structure of the result is:
\begin{equation}\label{eq:analytic_signal_model}
\begin{split}
    s^2_3(t, \theta) = & X_0 + X_1\cos{(\Delta \omega t + \Delta \varphi)} \\
    & + X_2\cos{(2\Delta \omega t + 2\Delta \varphi)} \\ 
    & + \cdots + X_{T}\cos{(T\Delta \omega t + T\Delta \varphi)},
\end{split}
\end{equation}
where $\Delta\omega$ and $\Delta\varphi$ are the beat frequency and the phase difference between the two sinusoids. The $X_k$ terms are complicated combinations of multiple powers (up to $2T$) the original amplitudes $A_1$ and $A_2$, the exponential decay terms and the $C_k$ coefficients.


This allows us to investigate how adding higher-order terms affects our approximation of $s_3(t)$. Here, we focus on cases where an odd number of terms is included in the model. A 1-term model would collapse to a single exponential fit, making it comparable to $M_{1}$.

\begin{figure}[H]

    \centering
    \includegraphics[width=1.0\linewidth]{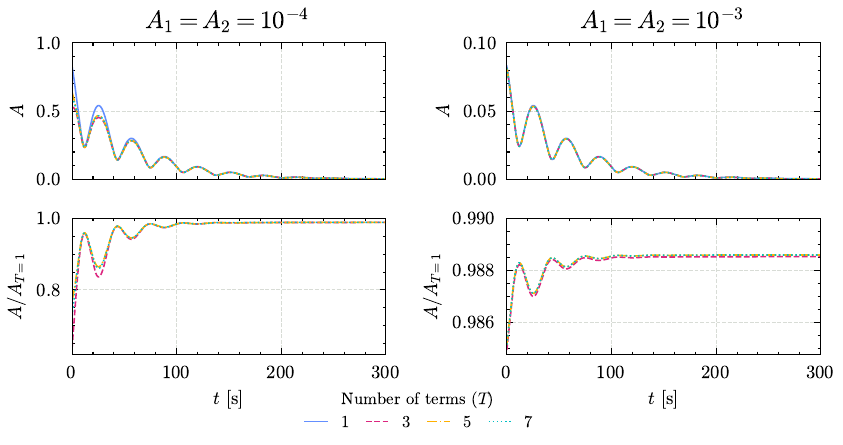}

    \caption{Analytic signal model with varying power series terms ($T$) and amplitudes. 
    \textbf{Upper} plots show the predicted amplitude versus time. \textbf{Lower} plots showing the ratio of amplitudes when including $T = {3, 5, 7}$ terms, compared to $T = 1$, for the two signals. These are compared with relatively larger amplitudes $A_1 = 10^{-3}\,\text{m}$, $A_2=10^{-3}\,\text{m}$ (left) and smaller amplitudes: $A_1 = 10^{-4}\,\text{m}$, $A_2 = 10^{-4}\,\text{m}$ (right). }
    \label{fig:examplesignals}
\end{figure}

In~\cref{fig:examplesignals}, we simulate an example signal using different modelling techniques while keeping $\omega_{n}$, $\tau_{n}$, and $\varphi_{n}$ constant. Since higher-order and non-linear contributions to the measured signal are expected to vary with amplitude, we consider both a high-amplitude case ($A_{1}=A_{2}=1 \times 10^{-3}\,\textrm{m}$) and a low-amplitude case ($A_{1}=A_{2}=1 \times 10^{-4}\,\textrm{m}$). The resulting ringdown measurements for models with 3, 5, and 7 terms are compared to the 1-term model, revealing significant differences of up to 25$\%$ in the high-amplitude case. It is also noted that in all cases, the models are able to better estimate the `troughs' of each oscillation, when there is lower overall motion of the laser spot. These discrepancies highlight the limitations of the previously used model described in \cref{models:model1}.

\section{Applying Bayesian Inference}\label{sec:bayes}

To compare the performance of the three models described in \cref{sec:models} we approach the problem using Bayesian inference.
Bayesian inference provides a probabilistic framework for inferring the parameters, $\vec{\theta}$, of some observed data, $d$, assuming an underlying model, $M$. At the core of this framework is Bayes’ theorem, which defines the probability of the parameters given the data and the model:
\begin{equation}\label{eq:bayes}
    p(\vec{\theta}|d, M) = \frac{p(d|\vec{\theta}, M)p(\vec{\theta}|M)}{p(d|M)},
\end{equation}
where $p(d|\vec{\theta}, M)$  is the likelihood, representing how well the observed data matches a given set of parameters under the model. The term  $p(\vec{\theta}|M)$  is the prior, which encodes any existing knowledge or assumptions about the parameters before considering the data. The denominator,  $p(d|M)$ , known as the Bayesian evidence (often denoted  $\mathcal{Z}$), acts as a normalization factor and is useful for comparing different models. A full review of this is given in \citeauthor{Sivia2006}~\cite{Sivia2006}.

Whilst this expression is simple to define, it is typically challenging to evaluate, and as a result, various techniques have been developed for performing Bayesian inference, such as Markov Chain Monte Carlo (MCMC) and Nested Sampling. This work employs the latter.
Bayesian inference also provides a framework evaluating how well different models describe the same, i.e. model comparison. To compare two models, one can compute the posterior odds between model 1 ($M_1$) and model 2 ($M_2$),
\begin{equation}\label{eq:posterior_odds}
    O_{12} = \frac{p(M_1)}{p(M_2)} \frac{p(d|M_1)}{p(d|M_2)}.
\end{equation}
The first term is the prior odds, this quantifies which model is preferred prior to observing the data, and the second term is the Bayes factor, $\mathcal{B}_{12}$; this quantified which model is preferred by the data. Typically, we express the Bayes factor in logarithmic scale for interpretability, then following \citeauthor{Jeffreys:1939xee}~\cite{Jeffreys:1939xee}: 
\begin{itemize}
    \item $\log_{10} \mathcal{B}_{12} \sim 0$ means neither model is favoured,
    \item  $\log_{10} \mathcal{B}_{12} > 1/2 \; (<-1/2)$ is substantial evidence $M_{1}$ ($M_{2}$),
    \item $\log_{10} \mathcal{B}_{12} > 1 \;(<-1)$ is strong evidence for $M_{1}$ ($M_{2}$),
    \item $\log_{10} \mathcal{B}_{12} > 3/2 \;(<-3/2)$ is very strong evidence for $M_{1}$ ($M_{2}$),
     \item $\log_{10} \mathcal{B}_{12} > 2 \;(<-2)$ is decisive for $M_{1}$ ($M_{2}$).
\end{itemize}
 Due to the high signal-to-noise ratio of our simulated data, reflecting the experiment,  the $\log_{10} \mathcal{B}_{12}$ values tend to be inflated. However, even with this effect, a larger value still robustly indicates a preferred model.

\subsection{Nested Sampling}\label{subsec:nested}

Nested Sampling is a technique for computing integrals over multidimensional parameter spaces, introduced in \citeauthor{Skilling2004}~\cite{Skilling2004, Skilling2006}. While it has various applications, its primary use is in Bayesian inference, where it computes the Bayesian evidence, the denominator in \cref{eq:bayes}, and produces samples from the posterior distribution as a by-product. The central idea of Nested Sampling is to transform a multidimensional integral into a one-dimensional integral, which can be approximated by evolving a set of samples. For a recent review of Nested Sampling in the physical sciences, see \citeauthor{Ashton:2022grj}~\cite{Ashton:2022grj}. 

 We choose Nested Sampling over methods such as MCMC since it systematically explores the parameter space starting from the prior making it particularly well-suited to multimodal problems, where the calculated probability distribution contains multiple distinct peaks. In our model, there are several degeneracies, which are scenarios where different sets of parameters are equally likely.  This is particularly prevalent in the decay parameters $\tau_{1}$, $\tau_{2}$. Nested Sampling efficiently handles these degeneracies, ensuring a thorough exploration of the parameter space and providing more reliable results. Furthermore, since it computes an estimate of the evidence, it enables us to compute the Bayes factor in \cref{eq:posterior_odds} and compare models. 


In this study, we use \nessai, a state-of-the-art Nested Sampling method that leverages machine learning to improve sampling efficiency~\cite{Williams_2021,Williams:2023ppp}. It is particularly well suited to problems where the likelihood is computationally expensive, since it can parallelize the calculation.
Furthermore, it includes a range of diagnostics that can be used to diagnose potential issues during sampling.

\subsection{Likelihood definition}

In order to perform Bayesian Inference, there are two key components that must be defined: the likelihood and the prior from \cref{eq:bayes}.
We start by defining a model for the data, since this will inform our choice of prior.

We assume the experimental data $\vec{d}$ is a vector of $N$ independent measurements taken at times $\vec{t}$. It is then modelled in terms of a time varying signal $s(\vec{t}, \vec{\theta})$, with parameters $\vec{\theta}$ , and two noise components $\vec{n}_\textrm{A}$ and $\vec{n}_\textrm{S}$. The first, $\vec{n}_\textrm{A}$ is stationary Gaussian noise (where the amplitude does not vary in time) and describes conventional noise sources present on the sampled voltage, such as electronic noise or digital-to-analogue-conversion (DAC) noise. The second, $\vec{n}_\textrm{A}$ is amplitude dependent noise that originates from sources present on the laser beam before photodetection and the subsequent subtraction of one side of the photodetector from the other, such as laser intensity noise (RIN). We assume this noise is also approximately Gaussian, however, in practice this approximation is only valid when the signal is above the noise floor.
With these taken into consideration we can then define the data as
\begin{equation}\label{eq:data_model}
    \vec{d} = s(\vec{t}, \theta) (1 + \vec{n}_\textrm{A}) + \vec{n}_\textrm{S}.
\end{equation}
Since we assume both noise components are Gaussian, the resulting distribution will also be Gaussian.
Specifically, if we assume both components have means of zero, so will the resulting distribution, and the variance of $i$'th sample will be
\begin{equation}
    \xi^2_i = s(t_i, \theta)^2 \xi_\textrm{A}^2  + \xi_\textrm{S}^2.
\end{equation}
where $\xi_\textrm{A}^2$ is the variance of $\vec{n}_A$ and $\xi_\textrm{S}^2$ is the variance of $\vec{n}_S$.
We can then define the likelihood as
\begin{equation}\label{eq:likelihood}
    p(\vec{d}|\vec{\theta}, M) = \prod_{i=1}^{N} \frac{1}{\sqrt{2 \pi \xi_i^{2}}} \exp\left\{-\frac{1}{2}\frac{[d_i - s(t_i, \vec{\theta)}]^2}{\xi_i^2}\right\}.
\end{equation}
We assume that $\xi_\textrm{A}$ and $\xi_\textrm{S}$ are unknown and must therefore also be inferred.
If $\xi_\textrm{A} = 0$, then this likelihood reduces to the stationary noise case used in \citeauthor{PhysRevLett.125.011102}~\cite{PhysRevLett.125.011102}.
The exact parameters $\vec{\theta}$ will depend on the signal model being used, however, in all cases we assume the priors over all parameters to be uniform. We also assume that, since each sample in the final data is the results of the FFT at a fixed sampling rate of 250\,ks, the noise is uncorrelated between samples in the final. \\ 

\noindent We consider the three signal models described in~\cref{sec:models}, \cref{eq:model_1_equation} (\textbf{$M_1$}), \cref{eq:model_2_equation} (\textbf{$M_2$}) and \cref{eq:model3_equation} (\textbf{$M_3$}).
These all share a set of common parameters that much be inferred, these are
\begin{itemize}
    \item {Two amplitudes $A_1$ and $A_2$, which when sampling, we re-parameterise using the ratio
    \begin{equation}
        \aratio = \frac{A_2}{A_1},
    \end{equation}
    and constrain $\aratio \leq 1$ to ensure $A_1 \geq A_2$, we then sample in $A_1$ and $\aratio$;
    }
    \item{two decay constants $\tau_1$ and $\tau_2$;}
    \item{the beat frequency $\Delta\omega$ which we constrain to be positive;}
    \item{the phase difference $\Delta\varphi$, defined as $\Delta\varphi = \varphi_1 - \varphi_2$,    %
    which we constrain to $[0, 2\pi]$.
    }
\end{itemize} 
These constraints are applied in order to remove degenerate solutions, e.g., two signals with $\rho = 0.5$ and $\rho=2$ and all the parameters the same are identical, which would increase the cost of performing inference.
In addition to these parameters, $M_{2}$ and $M_{3}$ have the parameters defined in the relevant equations and share an additional scaling parameter, $\ascale$, which we include to account for any scaling(s) from external gain's etc which are applied to a given ringdown measurement; such that
\begin{equation}\label{eq:model_with_scale}
    s_{\{2,3\}}(t_i, \vec{\theta}) = \ascale s_{\{2,3\}}(t_i, \vec{\theta}),
\end{equation}
further details about the inference can be found in \cref{app:inference_details}.

We implement the signal models and likelihood described previously in a Python package called \bayesbeat~\cite{michael_j_williams_2024_12804193}.
This also includes an interface to the Python implementation of \nessai~\cite{michael_j_williams_2024_10965503} with support for parallelising the likelihood computation.
The implementation makes use of Just-In-Time (JIT) compilation via \texttt{numba}~\cite{Lam:2015bfh} to accelerate the signal model computation.
\bayesbeat is available via PyPI (\url{https://pypi.org/project/bayesbeat/}) and the source code is available at~\cite{michael_j_williams_2024_12804193}.

\section{Validation using simulated data}
\label{sec:simulated_data}

To validate the proposed noise and signal models we begin by using simulated data. This allows for verification that the model parameters can be correctly inferred from a known state and allows us to understand any biases that may arise from, for example, differences in the models. 

\noindent The data is simulated using the full signal model (model 2, \cref{eq:model_2_equation}) with the relative intensity noise (RIN) of the laser and analogue to digital conversion (ADC) noise included. RIN is estimated at $10^{-3}$ based on the typical noise level of the unstabilised He-Ne used and has the effect of introducing amplidude dependant noise onto the readout signal. The ADC noise is $2.12\,V_\textrm{rms}/{\sqrt{\textrm{Hz}}}$ and is simply the digitisation noise limit associated with the hardware bit resolution and sample rates. The noise on the readout signal will be dominated by RIN when signals are large and will eventually run into ADC noise as the signal size decreases.
The process described in \cref{sec:experimental_setup} was emulated with random parameters to create a set of signals to test. 50 sets of data were generated with an \ac{FFT} taken every 0.2 seconds. Code to reproduce the data, the analyses and the results are included in the data release available at \url{http://bayesbeat.github.io/bayesbeat-paper}.

\subsection{Noise model validation}

We validate the proposed two-component noise model from \cref{eq:data_model} with the noise model used in previous studies,  which assumed purely stationary noise (i.e., $\xi_2 = 0$).
 Since our model accounts for amplitude-dependent noise, we incorporate the noise term $\xi_i$ when computing the residual $\mathcal{R}$ at time $t_i$:
\begin{equation}\label{eq:residuals}
\mathcal{R}i = \frac{d_i - s_i(t_i, \vec{\theta}^{*})}{\xi{i}}
\end{equation}
where $\vec{\theta}^{*}$ represents the maximum likelihood parameters obtained from the posterior distribution.\\ 

\begin{figure}
    \centering
    \includegraphics[width=1.0\linewidth]{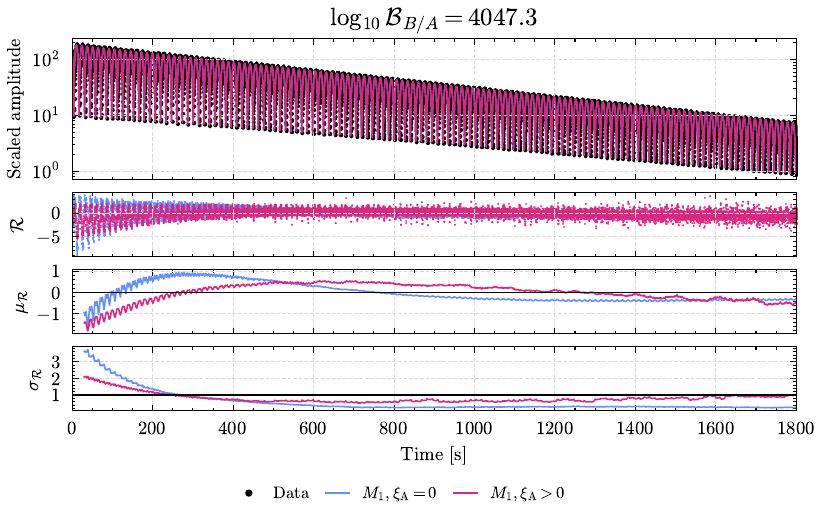}
    \caption{Fit and residuals for signal 31 obtained using the simple signal model (\textit{M$_{1}$}) with only stationary noise versus ($\xi_\textrm{A}=0$) stationary and amplitude dependent noise ($\xi_\textrm{A} < 0$) for a 2.5\,kHz mode (see \cref{fig:cylinder_modes}(a)). The log-Bayes factor is reported at the top of the figure. The lower two panels shows the mean $\mu_\mathcal{R}$ and standard deviation $\sigma_\mathcal{R}$ of the residuals as a function of time using a 30-second window. Since we use a Gaussian likelihood, if the signal fits the data perfectly, the mean should oscillate around zero and the standard deviation around one.}
    \label{fig:simple_model_fits_stationary}
\end{figure}

\noindent We analyse each of the 50 simulated ringdowns using the nested sampling algorithm described in \cref{sec:bayes}. This provides an estimate of the log-evidence and samples from the posterior distribution. We use the simple signal model ($M_1$) and repeat the analysis for the stationary noise model and the amplitude-dependent noise model. An example fit from this analysis is shown in \cref{fig:simple_model_fits_stationary}, data and code to produce the same plot for the other signals are available in the data release. This highlights how the residuals for the stationary noise exhibit a clear dependence on time, the rolling standard deviation decreases over time rather than remaining close to 1---the expected value since we have assumed a Gaussian likelihood. However, when using the amplitude-dependent noise model, the residuals do not show the same dependence. They do, however, show oscillations at larger amplitudes, these are present because the simple model does not include higher-order terms necessary to model these oscillations. The log-Bayes factor for this specific case is \logBFModelOneModelOneStationaryThirtyOne in favour of the updated noise model and serves to quantify the difference in the residuals. We also show the posterior samples for the decay parameters in beat frequency in \cref{fig:example_parameters}. These should that both analyses incorrectly infer the decay parameters. These results are discussed further in the following section once results for $M_3$ have been presented.

\begin{figure}
    \centering
    \includegraphics[width=0.7\linewidth]{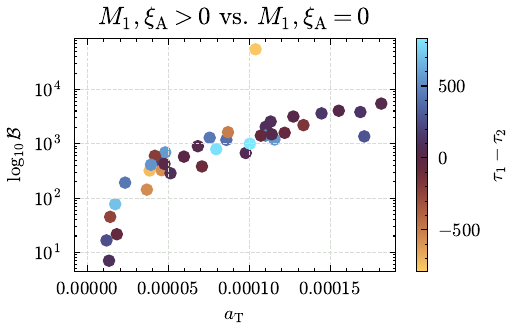}
    \caption{True total amplitude ($a_\textrm{T} = a_1 + a_2$) vs Bayes factor comparing analyses with a simple signal model ($M_1$) and two different noise hypotheses using simulated data; a noise model that includes stationary and amplitude dependant noise versus one that only includes stationary noise. The amplitudes quoted is the true amplitude for each signal. The colour of each point shows the different in the decay parameters.}
    \label{fig:noise_model_bf}
\end{figure}

To quantify which noise model is preferred, we compute the Bayes factor between the two models for 50 ringdowns and present the results in \cref{fig:noise_model_bf}. These results show the Bayes factor is dependent on the amplitude of the underlying signal---when the signal amplitude is low, the stationary noise is dominant and allowing $\xi_{A} < 0$ has minimal effects of the signal fit. We also see that the difference in the decay parameters, $\tau_1$ and $\tau_2$, also have little effect on the preferred noise model.

\subsection{Signal model validation}

Having validated the noise model, we now validate the proposed signal model. This allows us to verify that we can correctly infer the model parameters in simulated data and allows us to understand any biases that may arise when using different signal models.

\begin{figure}
    \centering
    \includegraphics[width=1.0\linewidth]{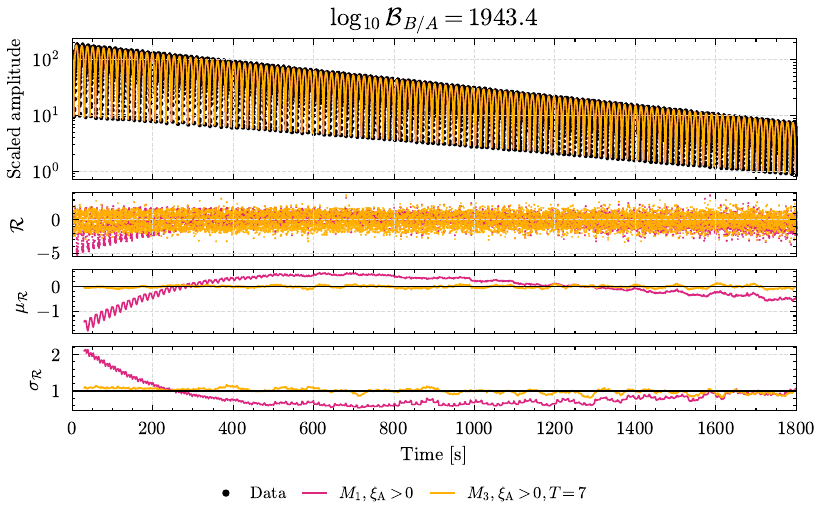}
    \caption{Fit and residuals for signal 31 obtained using the simple signal model (\textit{M$_{1}$}) with only stationary noise versus ($\xi_\textrm{A}=0$) stationary and amplitude dependent noise ($\xi_\textrm{A} >0$) for a 2.5\,kHz mode (see \cref{fig:cylinder_modes}(a)). The log-Bayes factor is reported at the top of the figure. The lower two panels shows the mean and standard deviation of the residuals as a function of time using a 30-second window. \Cref{fig:simple_model_fits_stationary} shows the results using the $M_1$ with the two different noise models.}
    \label{fig:simulated_fit_comparison}
\end{figure}

We consider the same signal as the previous section and compare the fit for the models 1 and 3, both with the amplitude dependent noise model ($\xi_A > 0$).
\Cref{fig:simulated_fit_comparison} shows both fits and their corresponding residuals. For the $M_3$, the residuals no longer show the same time-dependence and instead randomly fluctuate around a mean of zero and standard deviation of one. This confirms that the structure seen in the residuals for model 1 was indeed due to it lacking the higher-order terms. This is further supported by a log-Bayes factor of $\log_{10} \mathcal{B} = \logBFModelThreeModelOneThirtyOne$ in favour of model $M_3$ compared to $M_1$ , or equivalently, a log-Bayes factor of $\log_{10} \mathcal{B} = \logBFModelThreeModelOneStationaryThirtyOne$ for $M_3$ with amplitude-dependent noise versus $M_1$ with stationary noise.

\begin{figure}
    \centering
    \includegraphics[width=0.7\textwidth]{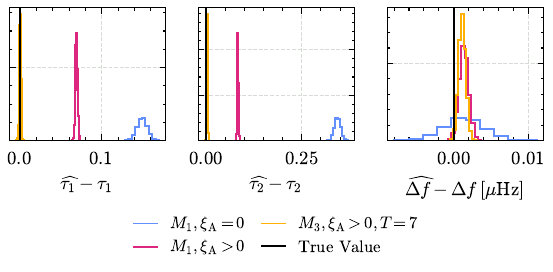}
    \caption{Comparison of the inferred posterior distributions (indicated with a hat) and true parameters for simulated signal 31. Results are shown for analyses using model 1 with stationary noise (blue), model 1 with amplitude-dependent noise and model 3 (with $T=7$) with amplitude-dependent noise. The true values have been subtracted from the posterior samples, such that the truth lies at zero.}
    \label{fig:example_parameters}
\end{figure}

We also compare the inferred parameters to the true parameters in \cref{fig:example_parameters}. This shows that, unlike model 1, model 2 can be used to accurately infer the decay parameters 

\begin{figure}
    \centering
    \includegraphics[width=0.7\textwidth]{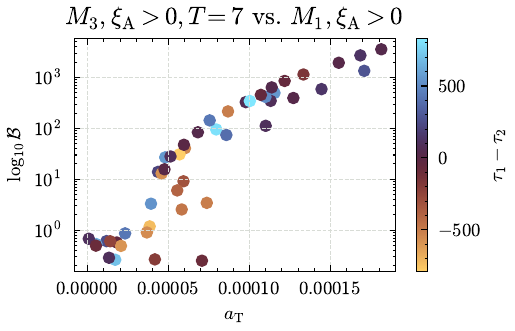}
    \caption{True total amplitude ($a_\textrm{T} = a_1 + a_2$) versus Bayes factors  comparing $M_{3}$ with $T=7$ vs $M_{1}$ for 50 simulated ringdowns. All analyses uses the noise model that include amplitude dependent noise.}
    \label{fig:simulated:bf_vs_a1}
\end{figure}

We compute the Bayes factor between the two models and present the results in \cref{fig:simulated:bf_vs_a1}.
These results show that $M_{3}$ is strongly favoured as the amplitude of the signal increases but at lower amplitudes neither model is favoured ($\log_{10}\mathcal{B} = 1$).
This is consistent with the effect shown in \cref{fig:Exponential_decay_distortion}, where we see the distortion is correlated with the signal amplitude.

\begin{figure}
    \centering
    \includegraphics[width=0.9\textwidth]{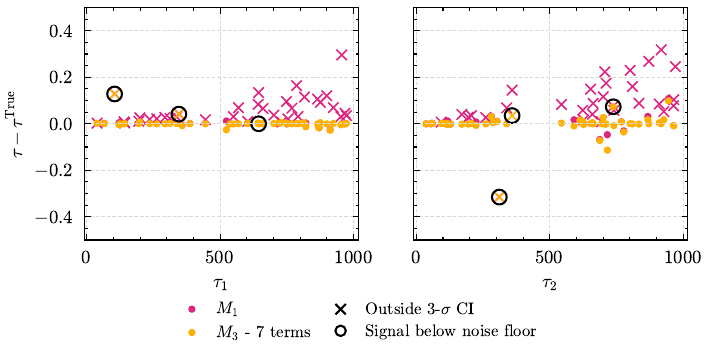}
    \caption{Absolute difference in inferred and true decay parameters for $M_1$ (blue) and $M_3$ with $T=7$ (orange when analysing 50 simulated ringdowns. The difference is computed using the median of the posterior distribution. Different markers are used to indicate whether the 3-$\sigma$ confidence interval contains the true value ($\circ$) or not ($\times$). Injections where the signal falls below the RIN noise floor are circled; note one omitted since the error is larger than 0.5, see \cref{app:model_3_limitations}.}
    \label{fig:tau_error}
\end{figure}

In \cref{fig:tau_error}, we show the difference  in inferred decay parameters compared to the true value for the injections. This shows the $M_3$ is consistently closer to the true value than $M_1$. Furthermore, $M_1$ often overestimates the decay parameters, with this being particularly apparent for larger values of $\tau_1$ --- this is also consistent with the results shown in \cref{fig:example_parameters}.
We also determine if the $95\%$ confidence interval for each posterior contains the true value and indicate this in \cref{fig:tau_error} using different markers.
For model 1, more than half of the results exclude the true value, suggesting the inference is often biased. In contrast, for $M_3$ only four ringdowns are marked as not including the true value in the 3-$\sigma$ confidence interval.
We examine these cases and find that they all share a common feature: the at some point in the data the signal falls below the noise floor due to RIN noise before applying the \ac{FFT}.
Since $M_3$ models the signal directly and does not account for the noise floor, it is unable to fit this feature which introduces a bias.
the biases are due to the signal fall below the noise floor in the initial 100 seconds of data -- an effect $M_3$ cannot capture since it does not include the noise floor. See \cref{app:model_3_limitations} for additional details.

\section{Results}\label{sec:results}

We characterise the mechanical loss of a sample using the ringdown technique described in~\cref{sec:experimental_setup}. In total 47 measurements were taken at frequencies between 2\,kHz\,-\,31\,kHz, a typical measurement range for samples of this geometry. For these runs, we set the beam size to  ($\sigma = 1.5 \, \text{mm}$), the size of the photodetector ($x_\textrm{e} = 10 \, \text{mm}$), and gap between the photodiodes to ($x_\textrm{g} = 0.25 \, \text{mm}$) based on measured values and compute the corresponding coefficients. The analyses took between 3 minutes to two hours to complete per-injection depending on the choice of model, more details about the cost are reported in \cref{app:analysis_cost}. Code to reproduce the analyses and the results from the analyses presented here are included in the data release available at \url{http://bayesbeat.github.io/bayesbeat-paper}.
 
\begin{figure}
    \centering
    \includegraphics[width=\linewidth]{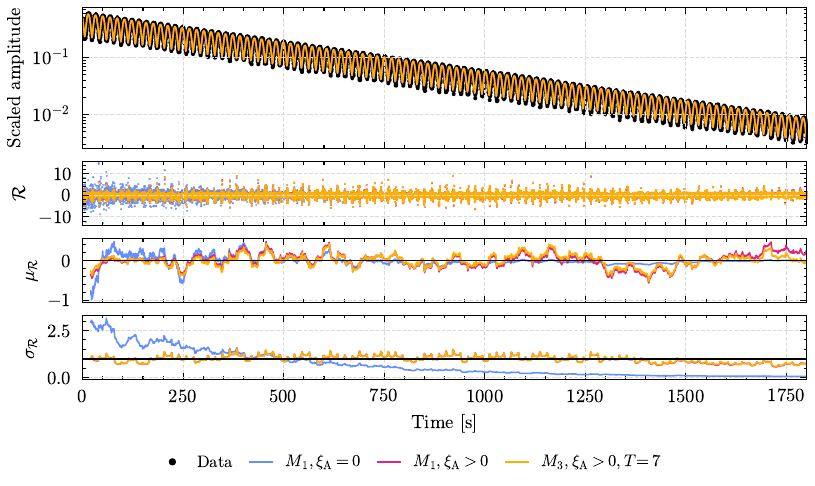}
    \caption{Fit and residuals for ringdown 0 obtained using $M_1$ with stationary noise (blue), $M_1$ with amplitude-dependent and stationary noise (orange) and $M_3$ with $T=7$ amplitude-dependent and stationary noise (green). The log Bayes factors are $\log_{10} \mathcal{B}=\logBFModelOneModelOneStationaryRealDataOne$ in favour of $M_1$ with both noise types compared to $M_1$ with only stationary noise and $\log_{10} \mathcal{B}=\logBFModelThreeModelOneRealDataOne$ in favour of $M_3$ compared to $M_1$, both including both noise types. The lower two panels shows the mean and standard deviation of the residuals as a function of time using a 30-second window.}
    \label{fig:real_data_fits}
\end{figure}

In \cref{fig:real_data_fits}, we show the fit and residuals for a ringdown at 2.5 kHz (see \cref{fig:cylinder_modes}(a)) analysed with the same three combinations of noise and signal models considered in \cref{sec:simulated_data}.

\begin{figure}
    \centering
    \includegraphics{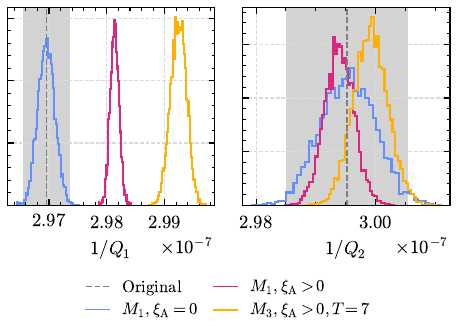}
    \caption{Posterior distributions for the mechanical loss ($1/Q$) computed from the inferred decay constants $\tau_1$ and $\tau_2$ for ringdown 0. Results are shown for two analyses with model 1, one with only stationary noise ($\xi_A=0$) and another with stationary and amplitude dependent noise ($\xi_A > 0$); and for an analysis with $M_{3}$ using $T = 7$ and including both types of noise. The results obtained using the original analysis from \citeauthor{Vajente_high_throughput}~\cite{Vajente_high_throughput} are shown in grey; the dashed line indicates the estimated value and the shaded region shows the estimated errors.}
    \label{fig:loss_comparison_0}
\end{figure}

Since we are interested in evaluating the effects of this new model on the estimation of the decay parameters, we also compare the inferred mechanical loss for the same ringdown and compare the results to those obtained with $M_{1}$ in \cref{fig:loss_comparison_0} and those obtained using the method described in \citeauthor{PhysRevLett.127.071101}~\cite{PhysRevLett.127.071101}.
As seen when analysing the simulated data, there is a clear difference in the inferred losses. The analyses with $M_1$ both underestimate the loss compared to the analysis with $M_3$.
This is consistent with this being a ringdown where $M_3$ is favoured over $M_1$, the log Bayes factor is \logBFModelThreeModelOneRealDataOne, and shows that the biases observed in \cref{sec:simulated_data} can also arise in real data.
It also shows that, for this ringdown, the results produced following \citeauthor{PhysRevLett.127.071101}~\cite{PhysRevLett.127.071101} are consistent with those inferred using nested sampling and $M_1$ with stationary noise.

\begin{figure}
    \begin{subfigure}[t]{0.48\textwidth}
        \centering
        \includegraphics[width=\textwidth]{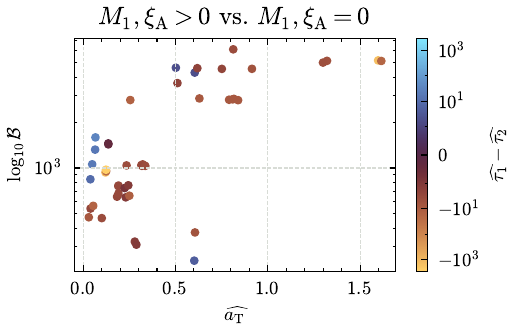}
        \caption{}\label{fig:bf_real_data_noise}
    \end{subfigure}
    \hfill
    \begin{subfigure}[t]{0.48\textwidth}
        \centering
        \includegraphics[width=\textwidth]{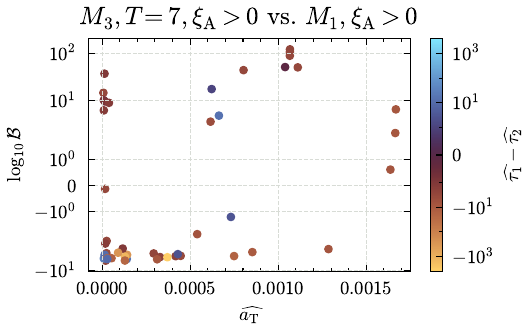}
        \caption{}
        \label{fig:bf_real_data_signal}
    \end{subfigure}
        \centering
        
    \caption{Inferred total amplitude with stationary noise, $\widehat{a_\textrm{T}} = a_1 + a_2$, versus Bayes factor comparing different models. The colour axis for each plot shows the difference between the inferred decay parameters.}
    \label{fig:bf_real_data}
\end{figure}

We report Bayes factors for all 47 ringdowns in \cref{fig:bf_real_data} as function of the inferred amplitudes, both different combinations of noise and signal models.
The results in \cref{fig:bf_real_data_noise} show the same behaviour as the results for simulated data in \cref{fig:bf_real_data_noise}; irrespective of the inferred total amplitude and difference in decay parameters, the analysis using $M_1$ and stationary and amplitude-dependent noise is heavily favoured compared to the analysis using $M_1$ and stationary noise with $\log_{10} \mathcal{B} > 100$.
\Cref{fig:bf_real_data_signal} shows that $M_3$ with $T=7$ is favoured for 17 of the 47 ringdowns, however, for the remainder either $M_1$ is favoured or neither is preferred.
Unlike for earlier results, there is only a rough correlation with amplitude. We examine the correlations with other parameters and find there is also a correlation with the amplitude ratio $\rho$, see \cref{fig:bf_real_data_a_ratio}, at lower total amplitudes $M_3$ is sometimes favoured in cases where the $\rho$ is between 0.1 and 0.5.
Cases where $M_1$ is favoured can be attributed to Occam's razor: if two models fit the data equally well, then the simpler model will be preferred and $M_3$ has more free parameters than $M_1$.

\begin{figure}
        \centering
        \includegraphics[width=\linewidth]{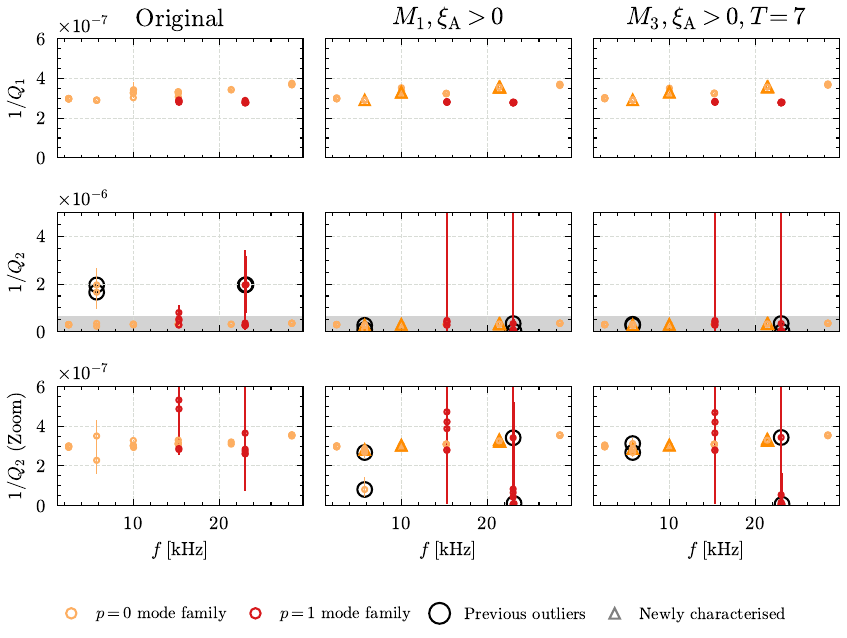}
    \caption{Posterior distributions of the mechanical loss inferred using $M_{1}$ with stationary noise and the original fitting method (left), $M_1$ both noise types and the proposed method (middle) and $M_{3}$ with $T=7$ and the proposed method (right) as a function of frequency. The error bars show the 95\% confidence regions. Points are coloured by their mode family. The original method can fail to produce errors and in these case the points are omitted and corresponding points for the proposed method are marked with a triangle ($\triangle$). The original method also predicts several outliers for $1/Q_2$, these are circled in black and the bottom panel only shows the shaded region of the $1/Q_2$ plot.}
    \label{fig:phi_results}
\end{figure}

\Cref{fig:phi_results} shows the losses inferred using the original method from \cite{Vajente_high_throughput} and the proposed method with two models: $M_1$ with both noise types and $M_{3}$ with $T=7$ and both noise types.
When comparing the primary loss $\tau_1$, we can see the results obtained using the improved models are far more tightly clustered and have smaller errors than the original results.
For the secondary loss, the most notable difference is the lack of significant outliers when using the new models; the original method could return spurious result but these no longer occur with our proposed approach. Similarly to the primary loss, the results are more consistent between measurements of the same modes, see e.g. the results for the $p=1$ mode family.
Furthermore, in the original analyses, there were six cases where the original method failed to produce valid fits, the majority of these were ringdowns that had reduced to single decaying sinusoids. The proposed method can, in contrast, fit such ringdowns without any adjustments, inferring posterior distributions for both $\tau_1$ and $\tau_2$. See \cref{app:additional_results} for an example of this.  Notably, in cases analysed with models $M_1$ and $M_3$, larger uncertainties are observed around frequencies of approximately 16 kHz and 23 kHz. These larger errors correspond to ringdowns with a high probability of being pure exponential decays—i.e., containing only one significant decay component ($\tau_1$). Previously, the original method misclassified these cases, leaving ambiguity regarding the interpretation of the fit results. Our Bayesian approach explicitly resolves this ambiguity, clearly distinguishing these pure decay cases.

\FloatBarrier
\section{Conclusions}

This work demonstrates a substantial advancement in analyzing ringdown measurements through an improved fitting methodology. By refining our modelling approach, we achieve significantly more accurate parameter extraction, directly overcoming the limitations of previous methods. Specifically, we establish that simpler Model 1 ($M_{1}$) with an amplitude dependant noise framework which adequately captures the dynamics at lower oscillation amplitudes. Whereas ringdowns with higher amplitudes, demand the complexity of Model 3 ($M_{3}$) to accurately represent the decay behaviour. This modelling strategy highlights the critical importance of matching model complexity to oscillation amplitude rather than applying uniform solutions. Additionally, this emphasizes the significant role of non-linear terms within the fitting models, which must be carefully accounted for to accurately represent system dynamics.

A critical finding of our analysis is the profound sensitivity of fitting outcomes to the chosen noise model and non-linear terms from the signal moving across the photodiode. Our results highlight the necessity of meticulously considering these characteristics, as different assumptions significantly affect parameter extraction. This also underscores the importance of the new proposed noise model introduced in this study, which represents a notable improvement over existing models. Whilst this model is a significant improvement over the previous method, it still has limitations. Most importantly, it does not account for the noise floor that results from applying the FFT. This suggests it could not be as accurate for signals that have decayed entirely below the noise floor since at that point the data will no longer be Gaussian. This will be a subject included in a later work.

These findings also highlight several limitations of the current experimental method, based on \ac{FFT} signal processing. This significantly increases the complexity of Bayesian analyses. Changing this processing  e.g. removing the \ac{FFT}, and heterodyning the signal could significantly simplify the analysis. Most importantly, this advanced fitting approach enables the analysis of modes that earlier methods deemed un-fittable with the previous method. By employing this more robust framework, we extract valuable insights from data previously considered unsuitable, significantly increasing out confidence in the estimation of the physical system. This increased accuracy in the estimation of $\tau_1$ and $\tau_2$ should in turn, increase our confidence in estimations of Coating Brownian thermal noise, and aid in our efforts to increase gravitational wave detector sensitivity. 

\ack
LIGO was
constructed by the California Institute of Technology and
Massachusetts Institute of Technology with funding from the
National Science Foundation, and operates under cooperative Agreement No. PHY-0757058. Advanced LIGO was built
under Award No. PHY-0823459. This paper carries LIGO
Document No. LIGO-P2500204.

The authors acknowledge Christopher Messenger, University of Glasgow, in aiding the development of the of our origional $M_{1}$ model. 

M.J.W. thanks Konstantin Leyde for insightful discussion about the likelihood definition.

Numerical computations were done on the Sciama High Performance Compute (HPC) cluster which is supported by the ICG, SEPNet and the University of Portsmouth.
We are grateful for computational resources provided by Cardiff University, and funded by STFC awards supporting UK Involvement in the Operation of Advanced LIGO.
This research is supported by the Science and Technology Facilities Council. J.B. and B.B. are supported by the Science and Technology Facilities Council (grant No. ST/V005634/1). M.J.W. acknowledges support from ST/X002225/1, ST/Y004876/1 and the University of Portsmouth.
This work made use of the following software packages: \texttt{corner}~\cite{corner}, \texttt{matplotlib}~\cite{Hunter:2007}, \texttt{nessai}~\cite{michael_j_williams_2024_10965503}, \texttt{numpy}~\cite{numpy}, \texttt{pandas}~\cite{reback2020pandas}, \texttt{sympy}~\cite{SymPy}

\appendix

\section{Inference details}\label{app:inference_details}

In this section, we provided additional details about the inference including details on the priors used and reparameterizations used during sampling. Code to reproduce all of the experiments is available at \url{http://bayesbeat.github.io/bayesbeat-paper}.

\subsection{Priors}\label{app:priors}

The cost of performing inference is directly related to the volume of the prior distribution being used to perform inference and the difference between this prior distribution and the posterior distribution. Previous analyses used an iterative approach to try and reduce the prior volume, however, we aim to avoid such approaches since they can introduce biases.

Instead, we only update the prior for the beat frequency ($\Delta \omega$) based on the observed data. This is achieved by applying a fourth-order low-pass butter filter, computing the \ac{FFT} of the data, and then finding any frequencies that have more power than the power at the low-pass frequency (0.01 Hz).
The prior distribution is then uniform, centred around this frequency with a user-specified width, $\delta$. We use $\delta=0.05\;\textrm{Hz}$ for real data and $\delta=1 / 10\pi\;\textrm{Hz}$ (0.2 rad) for simulated data.
In instances where no peak is found, we set the prior to $[0, \delta)$.

\subsection{Reparameterizations}\label{app:reparameterizations}

\begin{figure}
    \centering
    \includegraphics{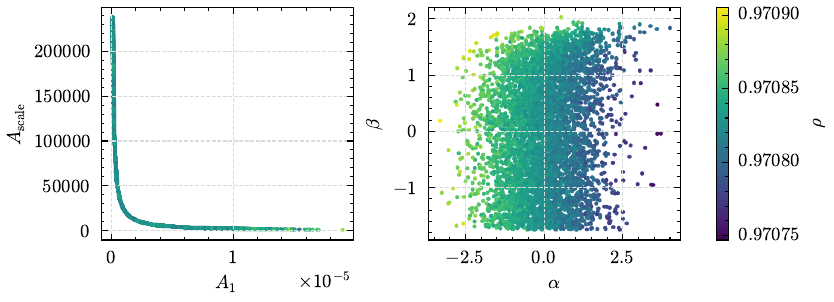}
    \caption{Example of applying the $\alpha$---$\beta$ reparameterization defined in \cref{eq:alpha_beta_reparameterization} to reparameterize $A_1$ and $\ascale$ posterior samples from the the analysis of the simulated ringdown 0 using $M_3$ with $T=7$ and both noise types. $\alpha$ and $\beta$ have been rescaled to have a mean of zero and standard deviation of 1. The colour axis shows the amplitude ratio $\rho = A_1 / A2$. }
    \label{fig:reparameterization}
\end{figure}

Inspired by a similar reparameterization used in \citeauthor{Veitch:2014wba}~\cite{Veitch:2014wba}, we use a linear combination of the log of $A_1$ and $A_\textrm{scale}$ to reduce improve sampling efficiency.
We define two new parameters:
\begin{subequations}\label{eq:alpha_beta_reparameterization}
    \begin{equation}
       \alpha = \log_{10} A_1 + \log_{10} \ascale,
    \end{equation}
    \begin{equation}
        \beta = \log_{10} A_1 - \log_{10} \ascale,
    \end{equation}
\end{subequations}
where $\ascale$ is the scale parameter defined in \cref{eq:model_with_scale}.
When implementing this in \bayesbeat and \nessai, we also include a second rescaling that ensures the samples have a mean of zero and a standard deviation of one, this improves the performance of the normalizing flow used in \nessai.
\Cref{fig:reparameterization} shows an example of applying this reparameterization to posterior samples from an analysis in \cref{sec:simulated_data}, this shows how $A_1$ and $\ascale$ are correlated and how using $\alpha$ and $\beta$ simplifies this.

\FloatBarrier
\section{Additional results}\label{app:additional_results}

\subsection{Model 3 limitations}\label{app:model_3_limitations}

\begin{figure}
    \centering
    \includegraphics[width=\linewidth]{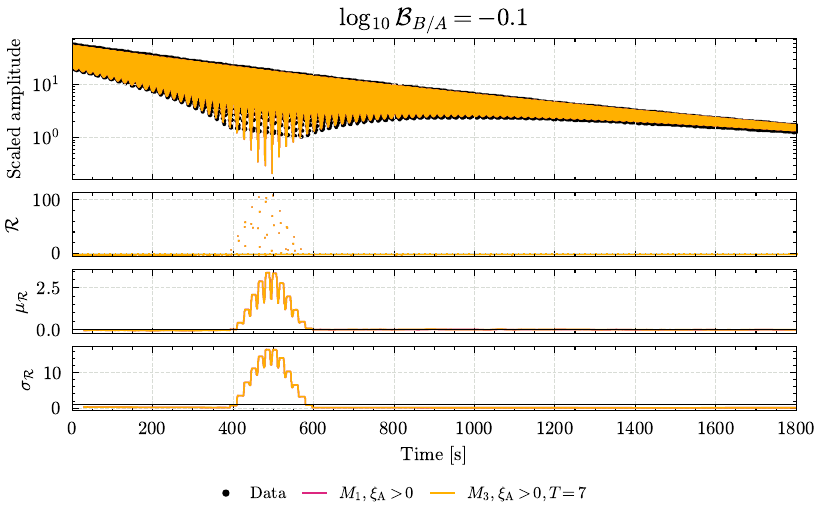}
    \caption{Fit and residuals for simulated ringdown 14 obtained using $M_1$ with amplitude-dependent and stationary noise (pink) and $M_3$ with $T=7$ amplitude-dependent and stationary noise (yellow). The log Bayes factor is $\log_{10} \mathcal{B}=\logBFModelThreeModelOneFourteen$ suggesting neither model is preferred. The lower two panels shows the mean and standard deviation of the residuals as a function of time using a 30-second window. This clearly shows where signal falls below the noise floor between 400 and 600 seconds.}
    \label{fig:model_3_noise_floor}
\end{figure}

When analyzing simulated data with $M_3$ in \cref{sec:simulated_data}, we found four ringdowns for which the inferred decay parameters were marginally biased. We inspect the data and find that all four share a common feature: at some point in the ringdown the signal amplitudes falls below the noise floor. \Cref{fig:model_3_noise_floor} shows an example of this. Since none of the models considered in this work include this feature, it introduces a bias in the fit and inferred parameters. We find that scale of these biases depends on the width of the priors, in most cases the bias is small ($<1\%$) but it one case we observe the bias is much larger since the inferred decay parameters are the swapped compare to the true values, i.e. the inferred $\tau_1$ is close the true $\tau_2$ value and vice versa. This effect can be avoided by making the priors narrower.

\begin{figure}
    \centering
    \includegraphics[width=0.9\linewidth]{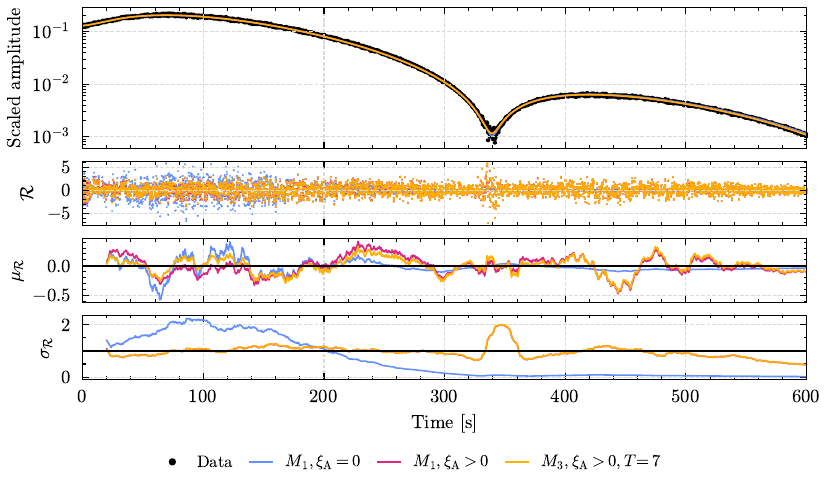}
    \caption{Fit and residuals for ringdown 43 obtained using $M_1$ with stationary noise (blue), $M_1$ with amplitude-dependent and stationary noise (orange) and $M_3$ with $T=7$ amplitude-dependent and stationary noise (green). The lower two panels shows the mean and standard deviation of the residuals as a function of time using a 30-second window. The residuals and rolling standard deviation clearly show how the signal amplitude falls below the noise floor between 300 and 400 seconds.}
    \label{fig:real_data_noise_floor}
\end{figure}

We examine the ringdowns analysed in \cref{sec:results}, and find several instances that show similar behaviour, such as for ringdown 43 which is shown in \cref{fig:real_data_noise_floor}.
In this case, the presence of the feature does not appear to significantly impact the inference, since the residuals for $M_3$ still have a mean close to zero and standard deviation close to one.

\subsection{Correlations between amplitude ratio and Bayes factor for real data}

\begin{figure}
    \centering
    \includegraphics[width=0.7\linewidth]{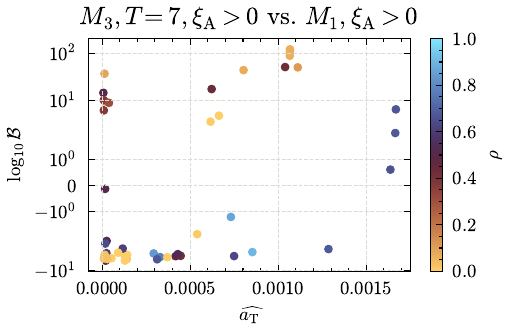}
    \caption{True total amplitude ($a_\textrm{T} = a_1 + a_2$) versus Bayes factors  comparing $M_{3}$ with $T=7$ vs $M_{1}$ for 47 ringdowns. All analyses uses the noise model that include amplitude dependent noise. The colourbar shows the amplitude ratio $\rho = a_2 / a_1$ and highlights how the Bayes factor is correlated with both total amplitude and $\rho$.}
    \label{fig:bf_real_data_a_ratio}
\end{figure}

As mentioned in \cref{sec:results}, when analyzing real data using the $M_3$ with $T=7$ and both noise models, we observe that preference for or against $M_3$ compared to $M_1$ is not only correlated with the total amplitude $a_\textrm{T}$ but also the amplitude ratio $\rho$. \Cref{fig:bf_real_data_a_ratio} is a modified version of \cref{fig:bf_real_data_signal} that includes $\rho$, it shows how for cases where the amplitudes differ by a factor of $\sim 2$, $M_3$ may be preferred even if the total amplitude is low.

\FloatBarrier
\subsection{Example fit to single-decay ringdown}

\begin{figure}
    \centering
    \includegraphics[width=\linewidth]{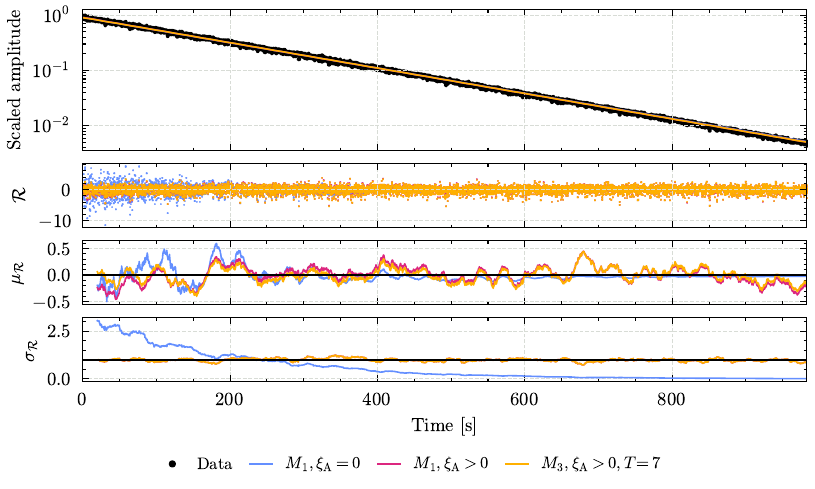}
    \caption{Fit and residuals for ringdown 4 obtained using $M_1$ with stationary noise (blue), $M_1$ with amplitude-dependent and stationary noise (orange) and $M_3$ with $T=7$ amplitude-dependent and stationary noise (green). The log Bayes factors are $\log_{10} \mathcal{B}=\logBFModelOneModelOneStationaryRealDataFour$ in favour of $M_1$ with both noise types compared to $M_1$ with only stationary noise and $\log_{10} \mathcal{B}= \logBFModelThreeModelOneRealDataFour$ in favour of $M_3$ compared to $M_1$, both including both noise types. The lower two panels shows the mean and standard deviation of the residuals as a function of time using a 30-second window.}
    \label{fig:fit_single_decay}
\end{figure}

\begin{figure}
    \centering
    \includegraphics{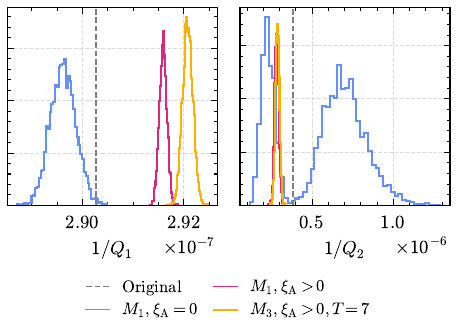}
    \caption{Posterior distributions for the mechanical loss ($1/Q$) computed from the inferred decay constants $\tau_1$ and $\tau_2$ for ringdown 4. Results are shown for two analyses with model 1, one with only stationary noise ($\xi_2=0$) and another with stationary and amplitude dependent noise ($\xi_A > 0$); and for an analysis with $M_{3}$ using $T = 3$ and including both types of noise. The results for the original analysis method from \citeauthor{PhysRevLett.127.071101}~\cite{PhysRevLett.127.071101}, are shown in grey; the dashed line indicates the estimated value and errors are omitted since the method fail to produce them.}
    \label{fig:loss_single_decay}
\end{figure}

In \cref{fig:fit_single_decay,fig:loss_single_decay}, we present results from the analysis of ringdown 4 from the data analysed in \cref{sec:results}.
This specific ringdown does not show a clear beat and instead resembles are single decaying sinusoid.
In our proposed framework, this requires no adjustments and the analyses can be started without having identified these features.
\Cref{fig:fit_single_decay} shows the fits for the three models considered in \cref{sec:results}. Once again, there is a clear preference for the models with both stationary and amplitude-dependent noise with a Bayes factor of $\log_{10} \mathcal{B} = \logBFModelOneModelOneStationaryRealDataFour$. However, in this case the additional complexity of $M_3$ is not required to effectively model the data and the Bayes factor for $M_1$ versus $M_3$ with $T=7$ is only $\log_{10}\mathcal{B} = \logBFModelThreeModelOneRealDataFour$ which is comparatively small in the context of data being analysed.

\FloatBarrier
\section{Analysis cost}\label{app:analysis_cost}

\begin{figure}
    \centering
    \includegraphics{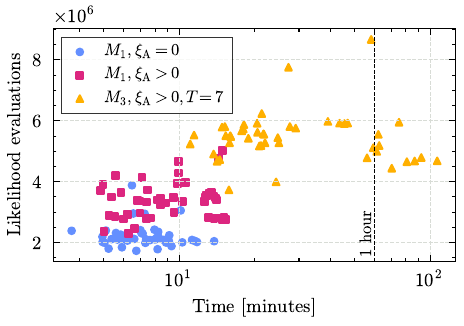}
    \caption{Total analysis time versus the total number of likelihood evaluations when analyzing the ringdowns described in \cref{sec:results} using $M_1$ with and without amplitude dependent noise and $M_3$ with amplitude dependent noise. Likelihood evaluations were parallelized over eight CPU cores.}
    \label{fig:inference_cost}
\end{figure}

\begin{figure}
    \centering
    \includegraphics{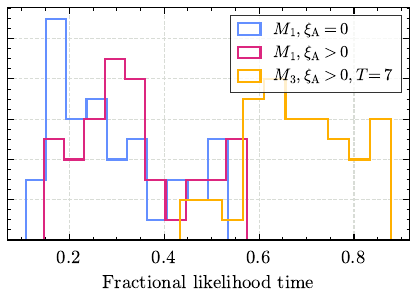}
    \caption{Fraction of the total analysis time spent evaluating the likelihood when analyzing the ringdowns described in \cref{sec:results} using $M_1$ with and without amplitude dependent noise and $M_3$ with amplitude dependent noise. Likelihood evaluations were parallelized over eight CPU cores.}
    \label{fig:ll_time_fraction}
\end{figure}

One downside to the more complete analyses employed in \cref{sec:results} are their cost compared to e.g. the method used in \citeauthor{PhysRevLett.127.071101}~\cite{PhysRevLett.127.071101}.
We examine the cost in terms of three quantities; the wall time as reported by Python, the total number of likelihood evaluations, and the fraction of time spent evaluating the likelihood. We present these statistics in \cref{fig:inference_cost,fig:ll_time_fraction}. All analyses were performed using eight cores to parallelize the likelihood calculation as implemented in \nessai.

When using the nested sampling the number of likelihood evaluations should scale with the number of parameters in the model. We see this behaviour in \cref{fig:inference_cost}: $M_1$ without amplitude dependent noise has seven parameters, $M_1$ with amplitude dependent noise has eight and $M_3$ with amplitude dependent noise has 10 and the number of likelihood evaluations increases accordingly.
The time taken for each analysis depends directly on the number of likelihood evaluations, the cost of each evaluation and the cost of the remaining steps in the sampling algorithm. \Cref{fig:inference_cost} shows that the sampling times for analyses with $M_1$ are comparable whilst the analyses with $M_3$ taken longer. Whilst this is in part due to the number of likelihood evaluation, $M_3$ is slower to evaluate than $M_1$ meaning that a large fraction of the total time is spent evaluating the likelihood. \Cref{fig:ll_time_fraction} show how this fraction compares more the different analyses and highlights that the analyses with $M_3$ are dominated by the cost of evaluating the likelihood. This suggests that the analyses could be accelerate by either reducing the cost of evaluating $M_3$ or increasing the number of CPU cores used for parallelization.

\FloatBarrier
\printbibliography

\end{document}